\documentclass[%
reprint,
superscriptaddress,
amsmath,amssymb,
aps,
]{revtex4-2}

\usepackage{graphicx}
\usepackage{dcolumn}
\usepackage{bm}
\usepackage{amsmath}
\usepackage{xcolor}
\usepackage[normalem]{ulem}

\begin{document}
	
	\preprint{APS/123-QED}
	
	\title{Interaction potentials for mutually induced dipoles in uniform fields}
	
	
	
	\author{Lucas H. P. Cunha}
	\email{lh1063@georgetown.edu}
	\affiliation{Institute for Soft Matter Synthesis and Metrology, Georgetown University, Washington, DC 20057}%
	
	
	\date{\today}
	
	\begin{abstract}
		Dipolar interactions govern the structure and dynamics of many soft-matter systems, from molecular assemblies to magnetic and polarizable colloids. When dipole moments are induced by an external field, mutual interactions lead to a many-body magnetization response that cannot be described by fixed-dipole models. Here, we derive the interaction potential for a system of mutually interacting induced dipoles in a uniform external field using a force-based approach. By accounting for the displacement-induced variation of the dipole moments, we obtain an interaction potential consisting of the classical dipole–dipole term supplemented by two- and three-body corrections arising from mutual induction. Comparisons with simplified models that neglect mutual magnetization reveal significant errors in the interaction potential, particularly in anisotropic particle assemblies. We also discuss an efficient $\mathcal{O}(N^2)$ iterative scheme for computing the mutual magnetization, enabling accurate simulations of large dipolar systems.
	\end{abstract}

	\maketitle
	
	\section{Introduction}
	
	Dipole-dipole interactions play a fundamental role in particle arrangements and dynamics at the molecular to colloidal scale. As a result, they are crucial to describe a variety of chemical, physical, and biological phenomena such as polymers solubility \cite{hughes2013, blanks1964}, colloids assembly \cite{halsey1990structure, spatafora2021, sherman2018, spatafora2024, tierno2014recent, cunha2024, du2017two}, suspensions rheology \cite{roure2018, rosa2019, morillas2020, de2011magnetorheological}, protein folding \cite{baldwin2007, ganesan2014, improta2008}, and magnetic resonance imaging \cite{them2017, wu2019, pinheiro2024}, to cite a few. The anisotropic nature of dipolar interactions is a crucial feature in shaping the particle arrangements into complex architectures, from chains to columns to network structures \cite{fermigier1992structure, furst2000dynamics, swan2014directed}. Moreover, the dynamic control over the dipoles' orientation by the application of time-varying fields is the basis for several promising advances in microtechnology, as the design of micro-robots \cite{dreyfus2005, babataheri2011, yang2017, zimmermann2022, spatafora2022, spatafora2023} and the development of tunable materials \cite{alharraq2020, ge2008, Sherman2019}. Still, the refined control necessary for these applications requires precise and efficient models to account for dipolar interactions.

	The dipolar moment might be an intrinsic property of the particle -- fixed dipole moment -- or induced by interactions with field sources. For instance, when certain particles are subject to an electric field, they develop a heterogeneous distribution of charges, resulting in polarization \cite{mittal2008polarization, dhont2010electric}. In the magnetic context, the initially uncorrelated magnetic moments inside the particles align with an applied magnetic field, resulting in the net magnetization of the particle. In weak fields, corresponding to the linear regime of the Langevin function used to model the magnetization of paramagnetic materials \cite{rosensweig2013}, the magnetization of a dipole reads $\textbf{m} = \chi \textbf{H}_{loc}$, where $\chi$ is the particle's magnetic susceptibility and $\textbf{H}_{loc}$ is the magnetic field at the particle's position. Similar relations are also valid in the electrostatic context.
	
	In a system of many particles, their induced fields disturb the magnetization of one another, creating a complex multi-body interaction response mathematically modeled by a linear system. This phenomenon is referred to in the literature as the mutual dipole magnetization/polarization \cite{spatafora2021, Sherman2019}, and plays a crucial role in the context of molecular and colloid polarization \cite{thole1981, jason2026}. As discussed by \citet{sherman2018}, ignoring dipolar mutual polarization in dispersions of colloids may lead not only to quantitative but also qualitative incorrect predictions for coexisting phases in equilibrium. More recently, \citet{jason2026} investigated how the mutual magnetization of paramagnetic particles is coupled to the internal structure and shape of clusters, and leads to anisotropic giant susceptibility in structured aggregates.
	
	Due to the sensitive response of the dipoles to the particles' arrangement, the free-energy in a system cannot be precisely computed using the classical relation for fixed dipoles. Instead, the appropriate relation follows from the counterpart of the free-energy in a magnetized continuous system $F = -1/2\int _V  \mu_{0}\textbf{M}\cdot \textbf{H}_{0} \, \text{d}V $ \cite{landau1884electrodynamics}, where $\mu_{0}$ is the magnetic permeability of the free space, $\textbf{M}$ is the local magnetization of the material, $\textbf{H}_0$ is the imposed external magnetic field, and $V$ is the volume. In a discrete system of $N$ particles, such relation reads $F = -1/2 \sum_i^N \mu_{0} \textbf{m}_i \cdot \textbf{H}_0$, where $\textbf{m}_i$ is the total magnetization of the particle $i$. 
	
	Here, we present a force-based derivation of the interacting potential for a discrete system of mutually induced paramagnetic particles under the action of a uniform external field. Starting from the magnetic force acting on a given particle, we develop the interacting potential carefully accounting for the disturbance caused by the displacement of the probed particle on the magnetization of the system. This approach complements standard field-theoretic derivations based on the work required to magnetize the system, and highlights the role of mutual induction in the resulting interaction potential. The remainder of the manuscript is organized as follows. First, we introduce the mutual magnetization in a system of interacting dipoles under the action of a magnetic field $\textbf{H}_0$, elaborating on the solution for the many-body problem. Then, we discuss the formulation for the potential of a single dipole in a non-uniform field, followed by the interacting potential for a pair in a uniform field, to introduce some of the mathematical concepts. Finally, we derive the general interacting potential for a system of $N$ particles. Lastly, we discuss the role played by the mutual dipole magnetization by comparing it with a model that ignores such mutual effects. 
	
	\begin{figure}
		\centering
		\includegraphics[width=.4\textwidth]{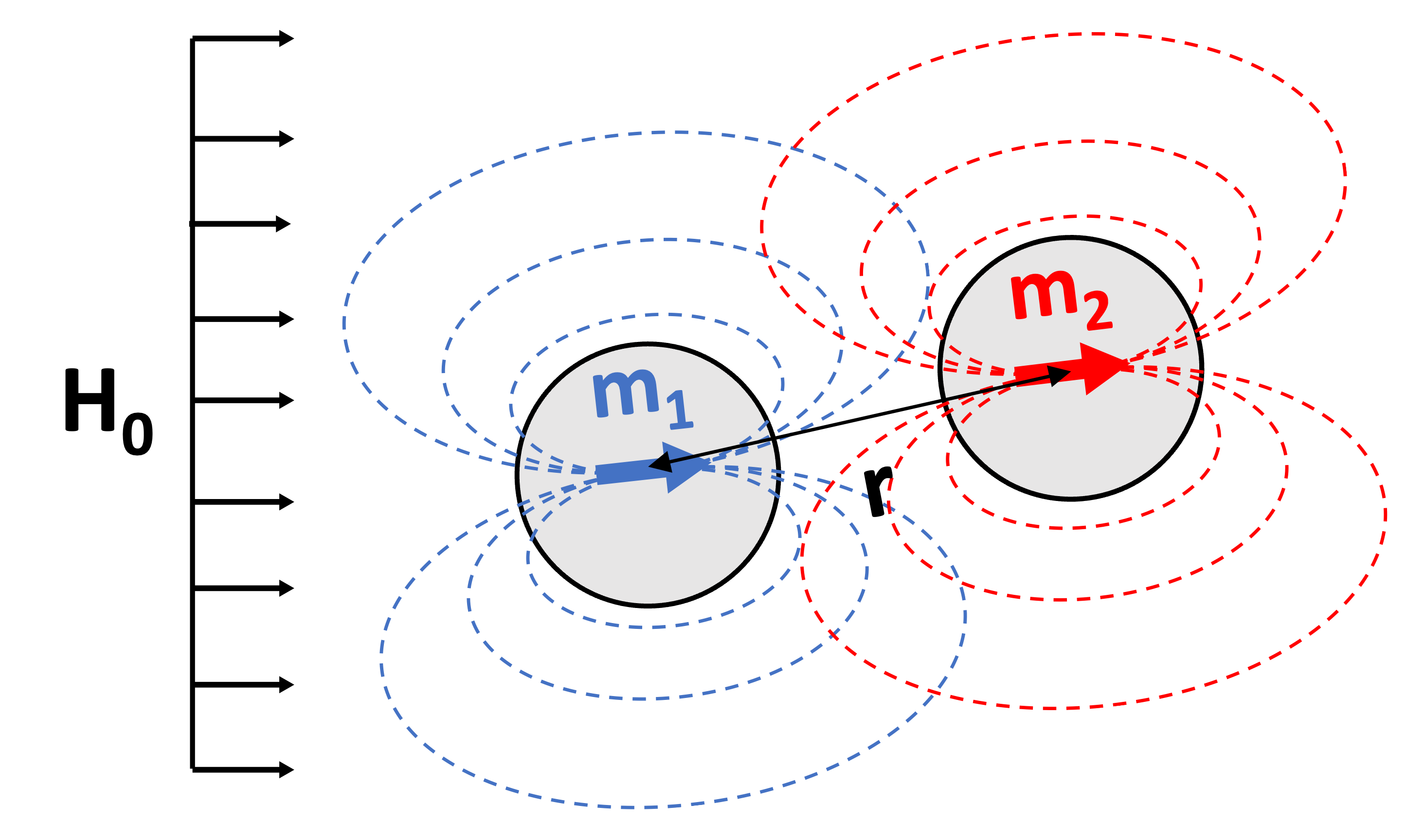}
		\caption{Sketch of a pair of mutually interacting magnetic dipoles separated by a distance $\textbf{r}$ subjected to an external magnetic field $\textbf{H}_0$. The dashed lines represent the field induced by each dipole.}
		\label{fig:sketch}
	\end{figure}
	
	\section{Solving for the magnetization of mutually interacting dipoles}
	\label{sec:mdm}
	The magnetization of a paramagnetic particle $i$ is given by $\textbf{m}_i =\chi_i\textbf{H}_{loc}(\textbf{x}_i)$,  where $\chi$ is the magnetic susceptibility of  the particle assumed as constant, $\textbf{x}_i$ is the position of particle $i$,  $\textbf{H}_{loc}(\textbf{x}_i) = \textbf{H}_0+\sum_{i\neq j}\textbf{H}^{ind}_{ij}$, $\textbf{H}_0$ is the external field, and $\textbf{H}_{ij}^{ind}$ is the field induced by particle $j$ at the position of particle $i$ \cite{spatafora2021}. From the point dipole approximation, we have 
	\begin{equation}
		\label{eq:H_ind}
		\textbf{H}^{ind}_{ij} = \mathcal{G}_{ij}\cdot \textbf{m}_j
	\end{equation}
	where 
	\begin{equation}
		\label{eq:G}
		\mathcal{G}_{ij} = \frac{1}{4\pi}\left[ 3 \frac{(\textbf{x}_i - \textbf{x}_j)(\textbf{x}_i - \textbf{x}_j)}{|(\textbf{x}_i - \textbf{x}_j)|^5} - \frac{\textbf{I}}{|(\textbf{x}_i - \textbf{x}_j)|^3} \right].
	\end{equation}
	\noindent
	Therefore, for a system of mutually interacting dipoles, we are led to
	\begin{equation}
		\label{eq:mdm_a}
		\textbf{m}_i = \chi_i \left( \textbf{H}_0 + \sum_{i\neq j} \mathcal{G}_{ij} \cdot \textbf{m}_j  \right ),
	\end{equation}
	\noindent
	which can be rewritten as
	\begin{equation}
		\label{eq:mdm_a2}
		\textbf{m}_i - \chi_i\sum_{i\neq j} \mathcal{G}_{ij}\cdot \textbf{m}_j = \chi_i\textbf{H}_0,
	\end{equation}
	\noindent
	or, in a matrix and vector form, as
	\begin{equation}
		\label{eq:linear}
		\begin{bmatrix}
			\mathbf{I}/\chi_1 & \mathcal{G}_{12} & \cdots  & \mathcal{G}_{1N}  \\
			\mathcal{G}_{21} & \mathbf{I}/\chi_2 & \cdots  & \mathcal{G}_{2N}  \\
			\vdots  &  & \ddots  &    \\
			\mathcal{G}_{N1} & \mathcal{G}_{N2} & \cdots  & \mathbf{I}/\chi_N   \\
		\end{bmatrix}
		\cdot
		\begin{bmatrix}
			\mathbf{m}_1 \\
			\mathbf{m}_2 \\
			\vdots \\
			\mathbf{m}_N
		\end{bmatrix}
		=
		\begin{bmatrix}
			\mathbf{H}_0 \\
			\mathbf{H}_0 \\
			\vdots \\
			\mathbf{H}_0
		\end{bmatrix} 
	\end{equation}
	\noindent
	where $N$ is the number of particles in the system, and $\textbf{I}$ is the $3\times 3$ unitary tensor.  The solution of the linear system in Eq.~\ref{eq:linear} gives the magnetization of all particles in the system. Classical methods to solve full-filed linear systems usually present a $\mathcal{O}(N^3)$ computational cost.
	
	Alternatively, one can solve the magnetization problem using an iterative method with $\mathcal{O}(N^2)$ computational cost, as follows. Equation~\ref{eq:mdm_a} may be rewritten in an recursive form as
	\begin{equation}
		\label{eq:recursive_1}
		\textbf{m}_i = \chi_i \sum_{C=0}^\infty \textbf{H}_i^{*(C)}
	\end{equation}
	\noindent
	for $\textbf{H}_i^{*(0)}=\textbf{H}_0$ and 
	\begin{equation}
		\label{eq:recursive_2}
		\textbf{H}_i^{*(C)} = \sum_i \sum_{j\neq i} \chi_j\mathcal{G}_{ij}\cdot \textbf{H}_j^{*(C-1)}.
	\end{equation}
	Important to note that the convergence of $\sum_{C=0}^\infty\textbf{H}_i^{*(C)}$ requires  $(\chi S/ a^{3})<1$, where $S$ is a geometrical factor \cite{jason2026} and $a$ corresponds to the radius of the particles and, so, it is the characteristic length of the problem for close packed systems.  
	
	\section{Pair of dipoles in a uniform field}
	
	We consider a pair of mutually interacting dipoles subjected to a uniform field $\textbf{H}_0(\textbf{x})=\textbf{H}_0$, as sketched in Fig. \ref{fig:sketch}. In this scenario, the mutual interaction between the particles' magnetization leads to 
	\begin{equation}
		\label{eq:mdm}
		\textbf{m}_1 = \chi_1\mathcal{G}_{2,1}(\textbf{r})\cdot\textbf{m}_2 + \chi_1 \textbf{H}_0,
	\end{equation}    
	\noindent 
	where $\textbf{r}$ is the distance vector between the two particles (see Fig.~\ref{fig:sketch}), and $\mathcal{G}_{1,2} = \mathcal{G}_{2,1}$. A similar expression is valid for the magnetization $\textbf{m}_2$, so that one can solve for the coupled magnetization of the dipoles, as elaborated in Sec.~\ref{sec:mdm}.  The force acting on the dipole $\textbf{m}_1$ due to the field generated by the dipole $\textbf{m}_2$  is then given by $\textbf{f}_{1} = \mu_0 \textbf{m}_1\cdot \nabla_{\textbf{x},1} \textbf{H}_{loc} = \mu_0 \textbf{m}_1\cdot \nabla_{\textbf{x},1} (\mathcal{G}_{2,1}\cdot \textbf{m}_2) $ . Here, $\nabla_{\textbf{x},i}$ refers to the spatial derivative evaluated at the position of particle $i$ without accounting for its or any other particle's displacement. This is an important detail in this problem due to the mutual magnetization between the two particles. Elaborating on the force acting on particle 1, we have
	\begin{eqnarray}
		\label{eq:force0}
		\frac{\textbf{f}_{1}}{\mu_0} &=& \textbf{m}_1\times (\nabla_{\textbf{x},1}\times\mathcal{G})\cdot\textbf{m}_2 + (\nabla_{\textbf{x},1}\mathcal{G}):\textbf{m}_2\textbf{m}_1 \nonumber \\ &=& (\nabla_{\textbf{r},1}\mathcal{G}):\textbf{m}_2\textbf{m}_1 \nonumber \\ &=& \nabla_{\textbf{r},1} (\mathcal{G}:\textbf{m}_2\textbf{m}_1) \nonumber \\ &&- (\nabla_{\textbf{r},1}\textbf{m}_1)\cdot \mathcal{G}\cdot\textbf{m}_2 - (\nabla_{\textbf{r},1}\textbf{m}_2)\cdot \mathcal{G}\cdot\textbf{m}_1 \nonumber \\
		&=& \nabla_{\textbf{r},1}\left[\mathcal{G}:\textbf{m}_2\textbf{m}_1 - \frac{1}{2}\chi_1(\mathcal{G}\cdot\textbf{m}_2)^2 - \frac{1}{2}\chi_2(\mathcal{G}\cdot\textbf{m}_1)^2 \right]  \nonumber \\ 
	\end{eqnarray}    
	\noindent
	where $\nabla_{\textbf{r},i}$ corresponds to the spatial differential operator regarding the displacement of the particle $i$. 
	In the development presented in Eq.~\ref{eq:force0} we use the facts that $\nabla_{\textbf{x},i}\times\mathcal{G}_{i,j} = \textbf{0}$, $\nabla_{\textbf{x},i}\mathcal{G}_{i,j}=\nabla_{\textbf{r},i}\mathcal{G}_{i,j}$, $\mathcal{G}_{i,j}=\mathcal{G}_{i,j}^T$, and $\textbf{m}_1 = \chi_1(\textbf{H}_0+\mathcal{G}_{2,1}\cdot \textbf{m}_2)$. Also, we dropped the particle indexes notion in $\mathcal{G}$ because $\mathcal{G}_{i,j}= \mathcal{G}_{j,i}$. One should note that we were careful  with the subtle difference between the vectorial operators $\nabla_\textbf{x}$ and $\nabla_{\textbf{r},i}$. If we develop the same analysis for particle 2, we end up with the same terms inside the gradient operator $\nabla_{\textbf{r},2}$. Thus, we have that the interacting potential for a pair of paramagnetic particles in a uniform field is given by
	\begin{equation}
		\label{eq:U7}
		U = -\mu_0\left[\mathcal{G} : \textbf{m}_1 \textbf{m}_2 - \frac{\chi_2}{2}(\mathcal{G}\cdot\textbf{m}_1)^2 - \frac{\chi_1}{2}(\mathcal{G}\cdot\textbf{m}_2)^2\right].
	\end{equation}
	\noindent
	The first term on the right-hand side of Eq.~\ref{eq:U7} corresponds to the classical potential for fixed dipoles, while the two other terms correspond to corrections due to the mutual polarization response to the particles' relative displacement. By adding the work corresponding to the magnetization of the isolated dipoles, $\mu_0(\chi_1 H_0^ 2/2 + \chi_2 H_0^ 2/2)$, and using Eq.~\ref{eq:mdm} to re-write the first term as $\mathcal{G}:[\chi_1(\mathcal{G}\cdot \textbf{m}_2 + \textbf{H}_0)\textbf{m}_2/2 + \chi_2(\mathcal{G}\cdot \textbf{m}_1 + \textbf{H}_0)\textbf{m}_1/2]$, we recover the relation for the free-energy in the system \cite{sherman2018, landau1884electrodynamics}
	\begin{equation}
		\label{eq:U7p}
		F = -\frac{\mu_0}{2}\left( \textbf{m}_1+\textbf{m}_2 \right)\cdot \textbf{H}_0.
	\end{equation}
	\noindent
	
	\section{3, 4, ..., N mutually interacting dipoles in a uniform field}
	\label{sec:many}
	
	The pair potential derived in Eq.~\ref{eq:U7} does not take into account systems with more than two particles and cannot be used to calculate the energy in the system as contributions from the interacting potentials between all pairs. For that, we must develop a potential that accounts for the many-body interactions. First, we derive the interacting potential for a system of three paramagnetic particles in a uniform magnetic field $\textbf{H}_0(\textbf{x}) = \textbf{H}_0$. Following a similar approach used for a pair, we elaborate on the force on particle 1.  
	\begin{equation}
		\label{eq:f1_1}
		\frac{\textbf{f}_1}{\mu_0} = \textbf{m}_1 \cdot \nabla_{\textbf{x},1} \left ( \textbf{H}_0 + \mathcal{G}_{1,2} \cdot \textbf{m}_2 + \mathcal{G}_{1,3} \cdot \textbf{m}_3   \right) 
	\end{equation}
	
	Given that $\nabla_\textbf{r} \textbf{m}_j = \chi_j \sum_{k\neq j} \nabla_\textbf{r}( \mathcal{G}_{jk} \cdot \textbf{m}_k)$ and $\textbf{H}_0$ is uniform, we get to
	
	\begin{eqnarray}
		\label{eq:f1_9}
		\frac{\textbf{f}_1}{\mu_0} & = &  \nabla_{\textbf{r},1}  [ \mathcal{G}_{1,2} : \textbf{m}_2  \textbf{m}_1 +  \mathcal{G}_{1,3} : \textbf{m}_3  \textbf{m}_1 +  \mathcal{G}_{2,3} : \textbf{m}_2  \textbf{m}_3  \nonumber \\
		& - &  \frac{\chi_2}{2}(\mathcal{G}_{1,2}\cdot \textbf{m}_{1})^2 - \frac{\chi_1}{2}(\mathcal{G}_{1,2}\cdot \textbf{m}_{2})^2 - \frac{\chi_3}{2}(\mathcal{G}_{1,3}\cdot \textbf{m}_{1})^2 \nonumber \\
		& - &  \frac{\chi_1}{2}(\mathcal{G}_{1,3}\cdot \textbf{m}_{3})^2 -\frac{\chi_2}{2}(\mathcal{G}_{2,3}\cdot \textbf{m}_{3})^2 - \frac{\chi_3}{2}(\mathcal{G}_{2,3}\cdot \textbf{m}_{2})^2  \nonumber \\
		& - &  \chi_1(\mathcal{G}_{1,3}\cdot \textbf{m}_3 \cdot \mathcal{G}_{1,2}\cdot \textbf{m}_2 ) 
		-  \chi_2(\mathcal{G}_{2,3}\cdot \textbf{m}_3 \cdot \mathcal{G}_{1,2}\cdot \textbf{m}_1 )   \nonumber \\
		& - &  \chi_3(\mathcal{G}_{2,3}\cdot \textbf{m}_2 \cdot \mathcal{G}_{1,3}\cdot \textbf{m}_1 ) ]
		-   (\nabla_{\textbf{r},1}\mathcal{G}_{2,3}):\textbf{m}_2\textbf{m}_3. \nonumber \\
	\end{eqnarray}
	\noindent 
	Note that $\nabla_{\textbf{r},i} \mathcal{G}_{j,k} = \mathbf{0}$ because the position of particle $i$ has no influence on $\mathcal{G}_{j,k}$, so the last term in the right-hand side of Eq.~\ref{eq:f1_9} is null. If we repeat the same analysis for the force on any of the other two particles, we get the same terms inside the $\nabla_{\textbf{r},i}$ operator. Thus, for three mutually interacting particles under a uniform magnetic field, the full interacting potential can be written as 
	\begin{eqnarray}
		\label{eq:U3}
		\frac{U}{\mu_0} & = &  -\mathcal{G}_{1,2} : \textbf{m}_2  \textbf{m}_1 -  \mathcal{G}_{1,3} : \textbf{m}_3  \textbf{m}_1 -  \mathcal{G}_{2,3} : \textbf{m}_2  \textbf{m}_3  \nonumber \\
		& + & \frac{\chi_1}{2}(\mathcal{G}_{1,2}\cdot \textbf{m}_{2})^2 + \frac{\chi_1}{2}(\mathcal{G}_{1,3}\cdot \textbf{m}_{3})^2 
		+ \frac{\chi_2}{2}(\mathcal{G}_{1,2}\cdot \textbf{m}_{1})^2
		\nonumber \\
		& + & \frac{\chi_2}{2}(\mathcal{G}_{2,3}\cdot \textbf{m}_{3})^2 +
		\frac{\chi_3}{2}(\mathcal{G}_{1,3}\cdot \textbf{m}_{1})^2 + \frac{\chi_3}{2}(\mathcal{G}_{2,3}\cdot \textbf{m}_{2})^2\nonumber \\
		& + &  \chi_1(\mathcal{G}_{1,3}\cdot \textbf{m}_3 \cdot \mathcal{G}_{1,2}\cdot \textbf{m}_2 ) +   \chi_2(\mathcal{G}_{2,3}\cdot \textbf{m}_3 \cdot \mathcal{G}_{1,2}\cdot \textbf{m}_1 )   \nonumber \\
		& + &  \chi_3(\mathcal{G}_{2,3}\cdot \textbf{m}_2 \cdot \mathcal{G}_{1,3}\cdot \textbf{m}_1 ).
	\end{eqnarray}
	\noindent
	Equation~\ref{eq:U3} brings a new three-body term $\chi_i( \mathcal{G}_{i,j}\cdot \textbf{m}_j)\cdot (\mathcal{G}_{i,k}\cdot \textbf{m}_k$), while the remaining are the same two-body terms from Eq.~\ref{eq:U7}.
	
	We repeat the same procedure for a system of four dipoles to get to
	\begin{eqnarray}
		\label{eq:4Un}
		\frac{U}{\mu_0} & = &  - \mathcal{G}_{1,2} : \textbf{m}_2  \textbf{m}_1 -  \mathcal{G}_{1,3} : \textbf{m}_3  \textbf{m}_1  -  \mathcal{G}_{1,4} : \textbf{m}_4  \textbf{m}_1  \nonumber \\
		& - &  \mathcal{G}_{2,3} : \textbf{m}_2  \textbf{m}_3 - \mathcal{G}_{2,4} : \textbf{m}_2  \textbf{m}_4 - \mathcal{G}_{3,4} : \textbf{m}_3  \textbf{m}_4  \nonumber \\
		& + &  \frac{\chi_2}{2}(\mathcal{G}_{1,2}\cdot \textbf{m}_{1})^2 + \frac{\chi_1}{2}(\mathcal{G}_{1,2}\cdot \textbf{m}_{2})^2 +
		\frac{\chi_3}{2}(\mathcal{G}_{1,3}\cdot \textbf{m}_{1})^2 \nonumber \\
		& + &  \frac{\chi_1}{2}(\mathcal{G}_{1,3}\cdot \textbf{m}_{3})^2 +
		\frac{\chi_4}{2}(\mathcal{G}_{1,4}\cdot \textbf{m}_{1})^2 + \frac{\chi_1}{2}(\mathcal{G}_{1,4}\cdot \textbf{m}_{4})^2 \nonumber \\
		& + &  \frac{\chi_2}{2}(\mathcal{G}_{2,3}\cdot \textbf{m}_{3})^2 + \frac{\chi_2}{2}(\mathcal{G}_{2,4}\cdot \textbf{m}_{4})^2
		+
		\frac{\chi_3}{2}(\mathcal{G}_{2,3}\cdot \textbf{m}_{2})^2 \nonumber \\
		& + &  \frac{\chi_3}{2}(\mathcal{G}_{3,4}\cdot \textbf{m}_{4})^2 
		+ \frac{\chi_4}{2}(\mathcal{G}_{2,4}\cdot \textbf{m}_{2})^2 + \frac{\chi_4}{2}(\mathcal{G}_{3,4}\cdot \textbf{m}_{3})^2  \nonumber \\
		& + & \chi_2 (\mathcal{G}_{2,3}\cdot \textbf{m}_{3} \cdot \mathcal{G}_{1,2} \cdot  \textbf{m}_1 ) + \chi_2(\mathcal{G}_{2,4}\cdot \textbf{m}_{4} \cdot \mathcal{G}_{1,2} \cdot  \textbf{m}_1) \nonumber \\  
		& + & \chi_1 (\mathcal{G}_{1,3}\cdot \textbf{m}_{3} \cdot \mathcal{G}_{1,2} \cdot  \textbf{m}_2 ) + \chi_1(\mathcal{G}_{1,4}\cdot \textbf{m}_{4} \cdot \mathcal{G}_{1,2} \cdot  \textbf{m}_2)  \nonumber \\  
		& + & \chi_3 (\mathcal{G}_{2,3}\cdot \textbf{m}_{2} \cdot \mathcal{G}_{1,3} \cdot  \textbf{m}_1 ) + \chi_3(\mathcal{G}_{3,4}\cdot \textbf{m}_{4} \cdot \mathcal{G}_{1,3} \cdot  \textbf{m}_1)  \nonumber \\
		& + & \chi_4 (\mathcal{G}_{2,4}\cdot \textbf{m}_{2} \cdot \mathcal{G}_{1,4} \cdot  \textbf{m}_1 ) + \chi_4(\mathcal{G}_{3,4}\cdot \textbf{m}_{3} \cdot \mathcal{G}_{1,4} \cdot  \textbf{m}_1) \nonumber \\
		& + & \chi_1 (\mathcal{G}_{1,3}\cdot \textbf{m}_{3} \cdot \mathcal{G}_{1,4} \cdot  \textbf{m}_4 ) + \chi_2 (\mathcal{G}_{2,3} \cdot\textbf{m}_3\cdot\mathcal{G}_{2,4} \cdot  \textbf{m}_4) \nonumber \\
		& + & \chi_3 (\mathcal{G}_{2,3} \cdot\textbf{m}_2\cdot\mathcal{G}_{3,4} \cdot  \textbf{m}_4) + \chi_4 (\mathcal{G}_{2,4} \cdot\textbf{m}_2\cdot\mathcal{G}_{3,4} \cdot  \textbf{m}_3). \nonumber \\
	\end{eqnarray}
	\noindent 
	
	Interestingly, the interacting potential for four dipoles presents no further type of term than those already shown for three dipoles, i.e., we get no four-body term. By extension, one has no reason to believe that the potential for five or more dipoles would require terms relating to more than three bodies, so we may extrapolate and generalize the potential for $N$ dipoles by
	\begin{eqnarray}
		\label{eq:many}
		\frac{U}{\mu_0} =  &-& \frac{1}{2}\sum_{i}^{N} \sum_{j\neq i}^{N}  \left[\mathcal{G}_{ij}:\textbf{m}_j \textbf{m}_i\right ] \nonumber \\
		& + & \frac{1}{2} \sum_{i}^{N} \sum_{j\neq i}^{N} \left[ \frac{\chi_j}{2}(\mathcal{G}_{ij}\cdot\textbf{m}_i)^2 + \frac{\chi_i}{2}(\mathcal{G}_{ij}\cdot\textbf{m}_j)^2\right] \nonumber \\ 
		& + & \frac{1}{2}\sum_{i}^{N} \sum_{j \neq i}^{N} \sum_{k\neq j,i}^{N} \left[ \chi_i (\mathcal{G}_{ij}\cdot\textbf{m}_j)\cdot(\mathcal{G}_{ik}\cdot\textbf{m}_k) \right] \nonumber \\
	\end{eqnarray}
	\noindent
	Again, the first term on the right-hand side in Eq.~\ref{eq:many} corresponds to the one for fixed-dipoles while the other two account for mutual dipolar corrections from two- and three-body interactions, respectively. 
	
	Using the relation $\textbf{m}_i = \chi_i(\textbf{H}_0 + \sum_{j\neq i} \mathcal{G}_{ij}\cdot\textbf{m}_j)$ in the first term on the right-hand side of Eq.~\ref{eq:many} and adding the work corresponding to the magnetization of the isolated dipoles $(-1/2\sum_i \mu_0\chi_i \textbf{H}_0^2)$ as done to get from Eq.~\ref{eq:U7} to Eq.~\ref{eq:U7p}, we are led to the discrete relation for the free-energy in the system  \cite{landau1884electrodynamics, jackson1998},
	\begin{equation}
		\label{eq:many2}
		F = -\frac{\mu_0}{2}\sum_i^N \textbf{m}_i \cdot \textbf{H}_0.
	\end{equation}
	
	\section{Analyzing the potential in different structures} \label{sec:results}
	In this section, we analyze the role played by mutual dipolar interactions (MDM) by comparing it with the simplified dipole model (DM), which neglects the mutual magnetization between particles. In the DM model, the particles' magnetization is given by $\textbf{m}_i = \chi_i \textbf{H}_0$ and the interacting potential is computed as $U_{DM} = -1/2\sum_i \sum_{j\neq i} \mathcal{G}_{ij}: \textbf{m}_i \textbf{m}_j$. The magnetic force on particle $i$ is calculated as $\textbf{f}_i = \sum_{j\neq i}(\nabla_{\textbf{r},i}\mathcal{G}_{ij}):\textbf{m}_i\textbf{m}_j$, as for the MDM case \cite{spatafora2021}. One should note that for the DM model, the interacting potential is calculated as for interacting fixed dipoles, even though we are dealing with magnetizable particles. Lastly, we compare the computational efficiency of using the expansion shown in Eqs.~\ref{eq:recursive_1} and~\ref{eq:recursive_2} with solving the linear system for the magnetization problem using classical solvers.

	\subsection{Pair of dipoles in a uniform field}
	\label{sec:an_pair}
\begin{figure} [h]
		\centering
		\includegraphics[width=.5\textwidth]{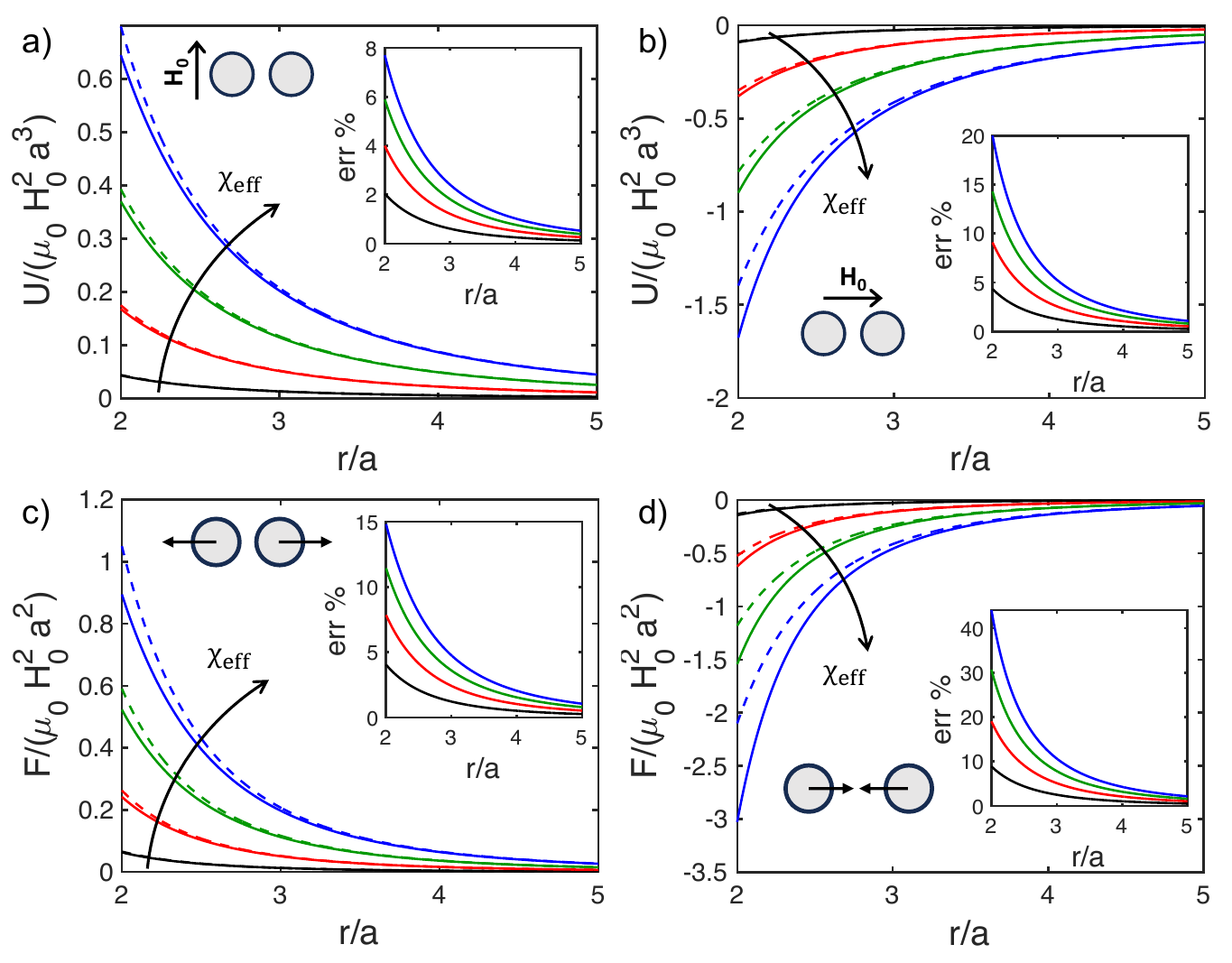}
		\caption{a) and b) present the pair potential as a function of $r$ for $\textbf{H}_0$ applied perpendicular to and parallel to $\textbf{r}$, respectively. c) and d) present the interparticle force as a function of $r$ for $\textbf{H}_0$ applied perpendicular to and parallel to $\textbf{r}$, respectively. Negative values correspond to attractive forces while positive corresponds to repulsive ones. The solid lines corresponds to the MDM model and the dashed lines corresponds to the DM model. Probed values of $\chi_{eff}$ are equal to 0.5 (black), 1.0 (red), 1.5 (green), and 2.0 (blue). The arrow perpendicular to the curves indicated increasing $\chi_{eff}$. Each graph presents an inset with the relative error in percentage of the DM model relative to the MDM model.}
		\label{fig:pot}
	\end{figure}

	Now, we analyze the classical problem of a pair of identical mutually interacting particles in a uniform magnetic field $\textbf{H}_0(\textbf{x})=\textbf{H}_0$. In the MDM case, $\textbf{m}_1=\chi(\textbf{H}_0 + \mathcal{G}_{2,1}\cdot\textbf{m}_2)$. For identical particles, the symmetry of the system leads to $\textbf{m}_1=\textbf{m}_2=\textbf{m}$, so \cite{thole1981, keaveny2008, cunha2024} 
	\begin{equation}
		\label{eq:mag}
		\textbf{m} = \chi\left( \textbf{I} - \chi\mathcal{G} \right)^{-1} \cdot \textbf{H}_0,
	\end{equation}
	\noindent
	and the pair potential in Eq.~\ref{eq:U7} becomes
	\begin{equation}
		\label{eq:Umag2}
		U = -\mu_0 \left[ \mathcal{G}:\textbf{m}\textbf{m} - \chi(\mathcal{G}\cdot\textbf{m})^2 \right].
	\end{equation}
	The magnetic susceptibility of the particles is given by $\chi = 4\pi a^3 \chi_{eff}/3$, where $a$ corresponds to their radii, $\chi_{eff} = 3\chi_m/(3+\chi_m)$ is the effective volumetric
	susceptibility for a sphere, and $\chi_m$ is the magnetic susceptibility of the material. For a realistic perspective in the context of paramagnetic colloidal particles, $\chi_{eff}$ is in the order of the unit \cite{cunha2024, spatafora2024}. The external field $\textbf{H}_0$ is either parallel or perpendicular to their relative distance $\textbf{r} = \textbf{x}_2 - \textbf{x}_1$.  Figures~\ref{fig:pot}a-d present the interacting potential and forces when considering the MDM and DM models as a function of $r$, for $\textbf{H}_0$ applied in the two relative directions and $\chi_{eff}$ varying from 0.5 to 2. 
	
	For $\textbf{H}_0$ applied perpendicular to $\textbf{r}$, the DM model overestimates both the interacting potential and the force between the particles. The relative errors increase for closer particles and higher $\chi_{eff}$, getting up to $\approx 8\%$ for the potential and $\approx 15\%$ for the force when $\chi_{eff}=2$ and $r/a=2$ (touching spheres). On the other hand, for $\textbf{H}_0$ applied parallel to $\textbf{r}$, the DM model underestimates the interacting potential and the force between the particles. As in the later case, the relative errors increase for closer particles and higher $\chi_{eff}$, getting up to $\approx 20\%$ for the potential and $\approx 40\%$ for the force when $\chi_{eff}=2$ and $r/a=2$ (touching spheres).
	
	\begin{figure}[h]
		\centering
		\includegraphics[width=.45\textwidth]{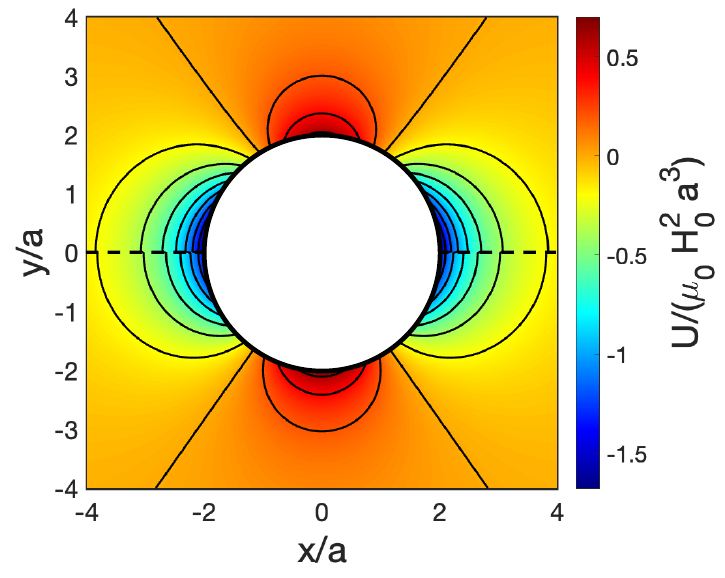}
		\caption{Pair potential for a probe paramagnetic particle placed at $\textbf{x} = x\,\hat{\textbf{e}}_x + y\,\hat{\textbf{e}}_y$ around another fixed at $\textbf{x} = \textbf{0}$, for $\textbf{H}_0 = H_0\,\hat{\textbf{e}}_x$ and $\chi_{eff}=2$. Above the dashed line $y/a>0$, the interacting potential corresponds to the MDM model. Below the dashed line $y/a<0$, the interacting potential corresponds to the DM model. The white region refers to the non-physical overlap between the two particles.}
		\label{fig:pot-field}
	\end{figure}
	
	Figure~\ref{fig:pot-field} presents the pair potential corresponding to having a probed paramagnetic particle at $\textbf{x} = x\,\hat{\textbf{e}}_x + y\,\hat{\textbf{e}}_y$, around another fixed at $\textbf{x} = \textbf{0}$, under the action of a uniform magnetic $\textbf{H}_0 = H_0 \hat{\textbf{e}}_x$, for $\chi_{eff}=2$. The upper portion of the figure, $y>0$, corresponds to the interacting potential for the MDM model, while the bottom portion corresponds to the DM model. The region contained in $\sqrt{x^2 + y^2}<2a$ is colored white because it represents the non-physical overlapping particles. The shapes of the two regions are mostly identical, meaning that using the DM model does not lead to dramatic qualitative errors for the configuration distributions for a dimer. The differences are only notable due to the discontinuity of the equipotential curves at $y=0$.

	\subsection{Cluster of particles in a uniform field}
	\label{sec:many}
\begin{figure}[h]
		\centering
		\includegraphics[width=0.5\textwidth]{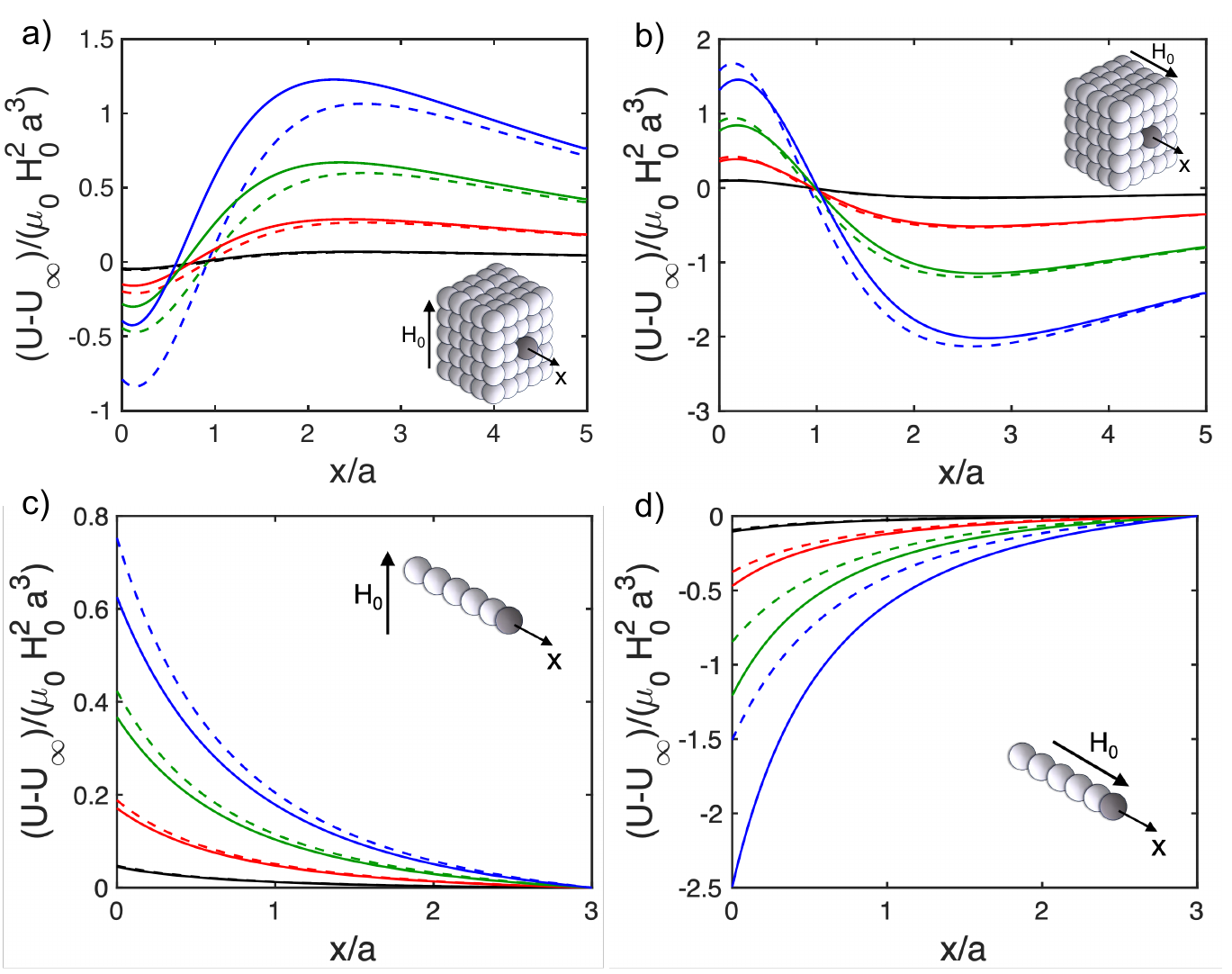}
		\caption{a) and b) total interacting potential for a cubic cluster of $5^3$ particles as a function of $\textbf{x}$ for $\textbf{H}_0$ applied perpendicular to and parallel to $\textbf{x}$, respectively. c) and d) total interacting potential for a linear chain of six particles as a function of $\textbf{x}$ for $\textbf{H}_0$ applied perpendicular to and parallel to $\textbf{x}$, respectively. The solid lines corresponds to the MDM model and the dashed lines corresponds to the DM model. Probed values of $\chi_{eff}$ are equal to 0.5 (black), 1.0 (red), 1.5 (green), and 2.0 (blue).}
		\label{fig:pot-cluster}
	\end{figure}

	As done in Sec.~\ref{sec:an_pair} for a pair of interacting dipoles, we study the role of mutual dipole interactions for clusters of many particles, focusing on anisotropic effects. For this, we analyze the interacting potential in a cubic cluster of $5^3$ particles (Fig.~\ref{fig:pot-cluster}ab) and in a linear chain of six particles (Fig.~\ref{fig:pot-cluster}cd). Initially, all neighboring particles are separated by a distance $2a$. The MDM and DM models are used to compute the interacting potential in the system, and the force on a probed particle as this is displaced a distance $\textbf{x}$ far from the cluster (see insets in Fig.~\ref{fig:pot-cluster}). $U_\infty$ represents the system's potential in the absence of the probe particle, i.e., $U_\infty = U(\textbf{x}\rightarrow \infty)$.
	
	For both the cubic and the chain clusters, we observe that the DM model overall well captures the behavior of the potential curve with respect to the displacement of the probed particle. However, we observe that the orientation of the field plays a major role in the deviation between the DM and the MDM calculations. For instance, the energy associated to the probed particle in the cubic cluster, $U(\textbf{x}=0)-U_{\infty}$, is overestimated by $\approx 105\%$ when $\textbf{H}_0$ perpendicular to $\textbf{x}$ and  overestimated by only $\approx 16\%$ when $\textbf{H}_0$ is parallel to $\textbf{x}$ for $\chi_{eff}=2$. For the chain structure, this discrepancy becomes even more drastic. The DM model overestimates the energy associated with the probed particle by $\approx20\%$ for $\textbf{H}_0$ perpendicular to the chain and underestimates the energy in $\approx 39\%$ for $\textbf{H}_0$ parallel to the chain when $\chi_{eff}=2$. Although the DM model can overall recover the behavior of the MDM model, its strong dependence on the field direction and the anisotropy of the structure may lead to poor quantitative and qualitative predictions for the field-induced assemblies and field-driven dynamics of magnetic colloids, as previously discussed by \citet{sherman2018}. In fact, the strong coupling between the particles' mutual magnetization and the geometry and the micro-structure of aggregates has been recently stressed by \citet{jason2026}.
	
	\subsection{$\mathcal{O}(N^2)$ approach for MDM}
	\label{sec:approx}
	Although it might be tempting to use the DM due to the computational costs to solve the magnetization problem, it is important to have in mind that this simplification may lead to quantitative and qualitative errors as discussed in Sec.~\ref{sec:many}. Alternatively, we here suggest an approach based on the expansion of Eq.~\ref{eq:recursive_1} with computational cost $\mathcal{O}(N^2)$, or cheaper when using cut-off schemes and neighboring lists for the simulation. For this, one should iterate Eq.~\ref{eq:recursive_2} until reaching an error $\mathrm{maxval}(|\textbf{H}^{*(C)} - \textbf{H}^{*(C-1)}|/|\textbf{H}_0^{*(C-1)}|$ smaller than a chosen tolerance, then using $\textbf{H}^{*(C)}$ into Eq.~\ref{eq:recursive_1}. 
	
\begin{figure}[h]
		\centering
		\includegraphics[width=0.5\textwidth]{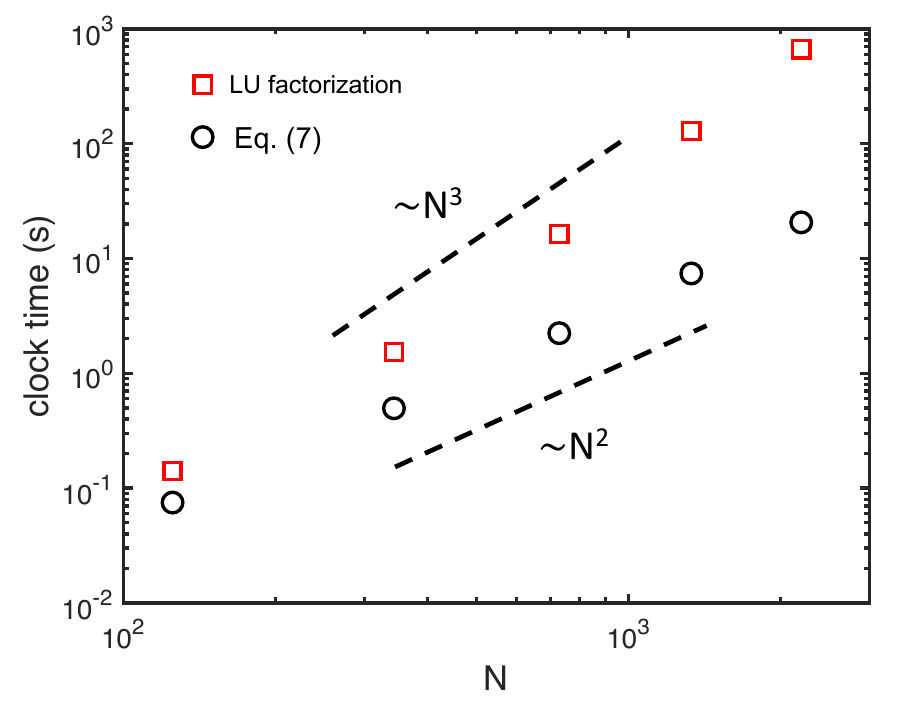}
		\caption{Computational time cost given in wall clock time solving the linear system by LU factorization and the iterative approach based on Eq.~\ref{eq:recursive_2} as a function of the number of particles $N$ in cubic clusters for $\chi_{eff}=2$. The dashed lines in black serve as a guide to the eyes.}
		\label{fig:time}
	\end{figure}

	Figure~\ref{fig:time} presents the clock time as a function of the number of particles $N$ in a cubic cluster to solve the magnetization problem by solving the linear system in Eq.~\ref{eq:linear} and iterating Eq.~\ref{eq:recursive_2} using a tolerance of $10^{-3}$. The linear system is solved using a LU factorization method. The computations are performed by a MATLAB code ran in a MacBook Pro with 18GB of memory and an Apple M3 Pro. As expected, we find a computational cost of $\mathcal{O}(N^2)$ for the iterative method while the classical LU factorization presents a cost of $\mathcal{O}(N^3)$. It is true that there are more efficient solvers than the classical LU factorization. But even the highly optimized inverted bar function of MATLAB ($A\setminus B$) presented a cost $\mathcal{O}(N^{2.33})$ for the given problem. In light of these results, we strongly recommend the use of the iterative method in Eq.~\ref{eq:recursive_2} for simulations of mutually interacting paramagnetic and polarizable particles.
	
	\section{Conclusion}
	We derived the interacting potential in a system of mutually interacting paramagnetic particles under the action of a uniform magnetic field from the force acting on the particles. The derived relation consists of the classical potential term for fixed dipoles and two further correcting terms to account for mutual magnetization from two- and three-body interactions. All calculations were developed in the magnetostatic context, but they also hold for polarizable particles subjected to electric fields. We demonstrate that ignoring mutual magnetization between particles, often done in the literature, leads to considerable errors in the interacting potential depending on the field direction and the system's geometry. Lastly, we elaborate on a computationally cheaper $\mathcal{O}(N^2)$ approach to solve for the mutual magnetization of the particles using an expansion of perpetuating induced fields. 
	
	\section*{Acknowledgments}
	I thank Dr. Mauro Mugnai, Dr. Aldo Spatafora-Salazar, and Prof. Peter Olmsted for the insightful discussions while working on the present problem. I am grateful for the support of the ISMSM-NIST Postdoctoral Fellowship at Georgetown University.
	
	\appendix
	
	\section{Potential for four mutually interacting dipoles}
	\label{sec:4dipoles}
	Here, we develop the potential for 4 mutually interacting dipoles under the action of a uniform external field. For this, we elaborate on the force acting on particle labeled as 1. Our goal is to include all terms inside de gradient operator. To make it easier for the reader visualization we underline all terms inside the gradient operator. 
	\begin{equation}
		\label{eq:4f1_1}
		\frac{\bm{f}_1}{\mu_0} = \bm{m}_1 \cdot \nabla_{x,1} \left ( \bm{H}_0 + \mathcal{G}_{1,2} \cdot \bm{m}_2 + \mathcal{G}_{1,3} \cdot \bm{m}_3 + \mathcal{G}_{1,4} \cdot \bm{m}_4   \right) 
	\end{equation}
	\noindent 
	Given that the external field is uniform, using the vectorial identity $\bm{a}\cdot\nabla\bm{b} = (\nabla\bm{b})\cdot\bm{a}-\bm{a}\times(\nabla \times \bm{b})$, and knowing that $\nabla_{x,i}\times(\mathcal{G}_{ij}\cdot\bm{m}_j) = \bm{0}$, we get to 
	\begin{eqnarray}
		\label{eq:4f1_2}
		\frac{\bm{f}_1}{\mu_0} & = &   \left (\nabla_{x,1} \mathcal{G}_{1,2} \right) : \bm{m}_2  \bm{m}_1 + \left (\nabla_{x,1} \mathcal{G}_{1,3} \right) : \bm{m}_3  \bm{m}_1 \nonumber \\
		& + & \left (\nabla_{x,1} \mathcal{G}_{1,4} \right) : \bm{m}_4  \bm{m}_1 
	\end{eqnarray}
	\noindent 
	As described in the main text, we have $\nabla_{x,i}\mathcal{G}_{ij} = \nabla_{r,i}\mathcal{G}_{ij}$ once the tensor $\mathcal{G}_{ij}$ is purely geometric.  Using the vectorial property $\nabla(\bm{a}\cdot\bm{b}) = (\nabla\bm{a})\cdot \bm{b} + (\nabla\bm{b})\cdot \bm{a}$, we get to
	\begin{eqnarray}
		\label{eq:4f1_4}
		\frac{\bm{f}_1}{\mu_0} & = &  \nabla_{r,1}  ( \underline{ \mathcal{G}_{1,2} : \bm{m}_2  \bm{m}_1 +  \mathcal{G}_{1,3} : \bm{m}_3  \bm{m}_1  +  \mathcal{G}_{1,4} : \bm{m}_4  \bm{m}_1 } ) \nonumber \\
		& - & (\nabla_{r,1}\bm{m}_2) \cdot \mathcal{G}_{1,2} \cdot  \bm{m}_1 -  (\nabla_{r,1}\bm{m}_1) \cdot \mathcal{G}_{1,2} \cdot  \bm{m}_2 \nonumber \\  
		& - & (\nabla_{r,1}\bm{m}_3) \cdot \mathcal{G}_{1,3} \cdot  \bm{m}_1 -  (\nabla_{r,1}\bm{m}_1) \cdot \mathcal{G}_{1,3} \cdot  \bm{m}_3 \nonumber \\
		& - & (\nabla_{r,1}\bm{m}_4) \cdot \mathcal{G}_{1,4} \cdot  \bm{m}_1 -  (\nabla_{r,1}\bm{m}_1) \cdot \mathcal{G}_{1,4} \cdot  \bm{m}_4 \nonumber \\
	\end{eqnarray}
	\noindent 
	Once $\bm{H}_0$ is uniform,  we have that $\nabla_r \bm{m}_j = \chi_j \sum_{k\neq j} \nabla_r( \mathcal{G}_{jk} \cdot \bm{m}_k)$. Then, we rewrite Eq.~\ref{eq:4f1_4} as
	\begin{eqnarray}
		\label{eq:4f1_5}
		\frac{\bm{f}_1}{\mu_0}  & = &  \nabla_{r,1}  ( \underline{ \mathcal{G}_{1,2} : \bm{m}_2  \bm{m}_1 +  \mathcal{G}_{1,3} : \bm{m}_3  \bm{m}_1  +  \mathcal{G}_{1,4} : \bm{m}_4  \bm{m}_1 }) \nonumber \\
		& - & \chi_2 \nabla_{r,1}(\mathcal{G}_{1,2}\cdot \bm{m}_{1} + \mathcal{G}_{2,3}\cdot \bm{m}_{3} + \mathcal{G}_{2,4}\cdot \bm{m}_{4}) \cdot \mathcal{G}_{1,2} \cdot  \bm{m}_1 \nonumber \\  
		& - & \chi_1 \nabla_{r,1}(\mathcal{G}_{1,2}\cdot \bm{m}_{2} + \mathcal{G}_{1,3}\cdot \bm{m}_{3} + \mathcal{G}_{1,4}\cdot \bm{m}_{4}) \cdot \mathcal{G}_{1,2} \cdot  \bm{m}_2 \nonumber \\  
		& - & \chi_3 \nabla_{r,1}(\mathcal{G}_{1,3}\cdot \bm{m}_{1} + \mathcal{G}_{2,3}\cdot \bm{m}_{2} + \mathcal{G}_{3,4}\cdot \bm{m}_{4}) \cdot \mathcal{G}_{1,3} \cdot  \bm{m}_1 \nonumber \\  
		& - & \chi_1 \nabla_{r,1}(\mathcal{G}_{1,2}\cdot \bm{m}_{2} + \mathcal{G}_{1,3}\cdot \bm{m}_{3} + \mathcal{G}_{1,4}\cdot \bm{m}_{4}) \cdot \mathcal{G}_{1,3} \cdot  \bm{m}_3 \nonumber \\  
		& - & \chi_4 \nabla_{r,1}(\mathcal{G}_{1,4}\cdot \bm{m}_{1} + \mathcal{G}_{2,4}\cdot \bm{m}_{2} + \mathcal{G}_{3,4}\cdot \bm{m}_{3}) \cdot \mathcal{G}_{1,4} \cdot  \bm{m}_1 \nonumber \\  
		& - & \chi_1 \nabla_{r,1}(\mathcal{G}_{1,2}\cdot \bm{m}_{2} + \mathcal{G}_{1,3}\cdot \bm{m}_{3} + \mathcal{G}_{1,4}\cdot \bm{m}_{4}) \cdot \mathcal{G}_{1,4} \cdot  \bm{m}_4 \nonumber \\  
	\end{eqnarray}
	\noindent 
	We now make the use of the vectorial identity $(\nabla\bm{a})\cdot\bm{a} = (1/2)\nabla(\bm{a}^2)$ to get to
	\begin{eqnarray}
		\label{eq:4f1_6}
		\frac{\bm{f}_1}{\mu_0} & = &  \nabla_{r,1}  [ \underline{ \mathcal{G}_{1,2} : \bm{m}_2  \bm{m}_1 +  \mathcal{G}_{1,3} : \bm{m}_3  \bm{m}_1  +  \mathcal{G}_{1,4} : \bm{m}_4  \bm{m}_1 } \nonumber \\
		& - & \underline{ \frac{\chi_2}{2}(\mathcal{G}_{1,2}\cdot \bm{m}_{1})^2 - \frac{\chi_1}{2}(\mathcal{G}_{1,2}\cdot \bm{m}_{2})^2 -\frac{\chi_3}{2}(\mathcal{G}_{1,3}\cdot \bm{m}_{1})^2 } \nonumber \\
		& - &  \underline{ \frac{\chi_1}{2}(\mathcal{G}_{1,3}\cdot \bm{m}_{3})^2 - \frac{\chi_4}{2}(\mathcal{G}_{1,4}\cdot \bm{m}_{1})^2 - \frac{\chi_1}{2}(\mathcal{G}_{1,4}\cdot \bm{m}_{4})^2 } ] \nonumber \\
		& - & \chi_2 \nabla_{r,1}(\mathcal{G}_{2,3}\cdot \bm{m}_{3} + \mathcal{G}_{2,4}\cdot \bm{m}_{4}) \cdot \mathcal{G}_{1,2} \cdot  \bm{m}_1 \nonumber \\  
		& - & \chi_1 \nabla_{r,1}(\mathcal{G}_{1,3}\cdot \bm{m}_{3} + \mathcal{G}_{1,4}\cdot \bm{m}_{4}) \cdot \mathcal{G}_{1,2} \cdot  \bm{m}_2 \nonumber \\  
		& - & \chi_3 \nabla_{r,1}(\mathcal{G}_{2,3}\cdot \bm{m}_{2} + \mathcal{G}_{3,4}\cdot \bm{m}_{4}) \cdot \mathcal{G}_{1,3} \cdot  \bm{m}_1 \nonumber \\  
		& - & \chi_1 \nabla_{r,1}(\mathcal{G}_{1,2}\cdot \bm{m}_{2} + \mathcal{G}_{1,4}\cdot \bm{m}_{4}) \cdot \mathcal{G}_{1,3} \cdot  \bm{m}_3 \nonumber \\  
		& - & \chi_4 \nabla_{r,1}(\mathcal{G}_{2,4}\cdot \bm{m}_{2} + \mathcal{G}_{3,4}\cdot \bm{m}_{3}) \cdot \mathcal{G}_{1,4} \cdot  \bm{m}_1 \nonumber \\  
		& - & \chi_1 \nabla_{r,1}(\mathcal{G}_{1,2}\cdot \bm{m}_{2} + \mathcal{G}_{1,3}\cdot \bm{m}_{3}) \cdot \mathcal{G}_{1,4} \cdot  \bm{m}_4 \nonumber \\  
	\end{eqnarray}
	\noindent 
	Making again the use the vectorial identity $\nabla{\bm{a}\cdot \bm{b}} = \nabla\bm{a} \cdot \bm  + \nabla\bm{b}\cdot \bm{a}$ to the appropriate terms, we can rewrite Eq.~\ref{eq:4f1_6} as
	\begin{eqnarray}
		\label{eq:4f1_8}
		\frac{\bm{f}_1}{\mu_0} & = &  \nabla_{r,1}  [ \underline{ \mathcal{G}_{1,2} : \bm{m}_2  \bm{m}_1 +  \mathcal{G}_{1,3} : \bm{m}_3  \bm{m}_1  +  \mathcal{G}_{1,4} : \bm{m}_4  \bm{m}_1 } \nonumber \\
		& - &  \underline{ \frac{\chi_2}{2}(\mathcal{G}_{1,2}\cdot \bm{m}_{1})^2 - \frac{\chi_1}{2}(\mathcal{G}_{1,2}\cdot \bm{m}_{2})^2 -\frac{\chi_3}{2}(\mathcal{G}_{1,3}\cdot \bm{m}_{1})^2 } \nonumber \\
		& - &  \underline{ \frac{\chi_1}{2}(\mathcal{G}_{1,3}\cdot \bm{m}_{3})^2 - \frac{\chi_4}{2}(\mathcal{G}_{1,4}\cdot \bm{m}_{1})^2 - \frac{\chi_1}{2}(\mathcal{G}_{1,4}\cdot \bm{m}_{4})^2 } \nonumber \\
		& - & \underline{ \chi_2 (\mathcal{G}_{2,3}\cdot \bm{m}_{3} \cdot \mathcal{G}_{1,2} \cdot  \bm{m}_1 ) - \chi_2(\mathcal{G}_{2,4}\cdot \bm{m}_{4} \cdot \mathcal{G}_{1,2} \cdot  \bm{m}_1) } \nonumber \\  
		& - & \underline{ \chi_1 (\mathcal{G}_{1,3}\cdot \bm{m}_{3} \cdot \mathcal{G}_{1,2} \cdot  \bm{m}_2 ) - \chi_1(\mathcal{G}_{1,4}\cdot \bm{m}_{4} \cdot \mathcal{G}_{1,2} \cdot  \bm{m}_2) } \nonumber \\  
		& - & \underline{ \chi_3 (\mathcal{G}_{2,3}\cdot \bm{m}_{2} \cdot \mathcal{G}_{1,3} \cdot  \bm{m}_1 ) - \chi_3(\mathcal{G}_{3,4}\cdot \bm{m}_{4} \cdot \mathcal{G}_{1,3} \cdot  \bm{m}_1) }  \nonumber \\
		& - & \underline{ \chi_4 (\mathcal{G}_{2,4}\cdot \bm{m}_{2} \cdot \mathcal{G}_{1,4} \cdot  \bm{m}_1 ) - \chi_4(\mathcal{G}_{3,4}\cdot \bm{m}_{3} \cdot \mathcal{G}_{1,4} \cdot  \bm{m}_1) } \nonumber \\
		& - & \underline{ \chi_1 (\mathcal{G}_{1,3}\cdot \bm{m}_{3} \cdot \mathcal{G}_{1,4} \cdot  \bm{m}_4 ) } ] \nonumber \\
		& + & \chi_2 \nabla_{r,1}(\mathcal{G}_{1,2} \cdot  \bm{m}_1)  \cdot (\mathcal{G}_{2,3}\cdot \bm{m}_{3} + \mathcal{G}_{2,4}\cdot \bm{m}_{4}) \nonumber \\  
		& + & \chi_3 \nabla_{r,1}(\mathcal{G}_{1,3} \cdot  \bm{m}_1)  \cdot (\mathcal{G}_{2,3}\cdot \bm{m}_{2} + \mathcal{G}_{3,4}\cdot \bm{m}_{4}) \nonumber \\    
		& + & \chi_4 \nabla_{r,1}(\mathcal{G}_{1,4} \cdot  \bm{m}_1)  \cdot (\mathcal{G}_{2,4}\cdot \bm{m}_{2} + \mathcal{G}_{3,4}\cdot \bm{m}_{3}) \nonumber \\  
	\end{eqnarray} 
	\newpage
	
	Notably, we have $\chi_2\nabla_{r,1}(\mathcal{G}_{1,2}\cdot\bm{m}_1) = (\nabla_{r,1}\bm{m}_2) - \nabla_{r,1}(\mathcal{G}_{2,3}\cdot\bm{m}_3 + \mathcal{G}_{2,4}\cdot\bm{m}_4)$, and so on. Then, 
	
	\begin{eqnarray}
		\label{eq:4f1_9}
		\frac{\bm{f}_1}{\mu_0} & = &  \nabla_{r,1}  [ \underline{ \mathcal{G}_{1,2} : \bm{m}_2  \bm{m}_1 +  \mathcal{G}_{1,3} : \bm{m}_3  \bm{m}_1  +  \mathcal{G}_{1,4} : \bm{m}_4  \bm{m}_1  }\nonumber \\
		& - &  \underline{ \frac{\chi_2}{2}(\mathcal{G}_{1,2}\cdot \bm{m}_{1})^2 - \frac{\chi_1}{2}(\mathcal{G}_{1,2}\cdot \bm{m}_{2})^2 -\frac{\chi_3}{2}(\mathcal{G}_{1,3}\cdot \bm{m}_{1})^2 } \nonumber \\
		& - & \underline{ \frac{\chi_1}{2}(\mathcal{G}_{1,3}\cdot \bm{m}_{3})^2 - \frac{\chi_4}{2}(\mathcal{G}_{1,4}\cdot \bm{m}_{1})^2 - \frac{\chi_1}{2}(\mathcal{G}_{1,4}\cdot \bm{m}_{4})^2 } \nonumber \\
		& - & \underline{ \chi_2 (\mathcal{G}_{2,3}\cdot \bm{m}_{3} \cdot \mathcal{G}_{1,2} \cdot  \bm{m}_1 ) - \chi_2(\mathcal{G}_{2,4}\cdot \bm{m}_{4} \cdot \mathcal{G}_{1,2} \cdot  \bm{m}_1) } \nonumber \\  
		& - & \underline{ \chi_1 (\mathcal{G}_{1,3}\cdot \bm{m}_{3} \cdot \mathcal{G}_{1,2} \cdot  \bm{m}_2 ) - \chi_1(\mathcal{G}_{1,4}\cdot \bm{m}_{4} \cdot \mathcal{G}_{1,2} \cdot  \bm{m}_2) } \nonumber \\  
		& - & \underline{ \chi_3 (\mathcal{G}_{2,3}\cdot \bm{m}_{2} \cdot \mathcal{G}_{1,3} \cdot  \bm{m}_1 ) - \chi_3(\mathcal{G}_{3,4}\cdot \bm{m}_{4} \cdot \mathcal{G}_{1,3} \cdot  \bm{m}_1)  } \nonumber \\
		& - & \underline{ \chi_4 (\mathcal{G}_{2,4}\cdot \bm{m}_{2} \cdot \mathcal{G}_{1,4} \cdot  \bm{m}_1 ) - \chi_4(\mathcal{G}_{3,4}\cdot \bm{m}_{3} \cdot \mathcal{G}_{1,4} \cdot  \bm{m}_1) } \nonumber \\
		& - & \underline{ \chi_1 (\mathcal{G}_{1,3}\cdot \bm{m}_{3} \cdot \mathcal{G}_{1,4} \cdot  \bm{m}_4 ) }] \nonumber \\
		& + & (\nabla_{r,1}\bm{m}_2)  \cdot (\mathcal{G}_{2,3}\cdot \bm{m}_{3} + \mathcal{G}_{2,4}\cdot \bm{m}_{4}) \nonumber \\  
		& - & \chi_2 \nabla_{r,1}(\mathcal{G}_{2,3} \cdot  \bm{m}_3 + \mathcal{G}_{2,4} \cdot  \bm{m}_4)  \cdot (\mathcal{G}_{2,3}\cdot \bm{m}_{3} + \mathcal{G}_{2,4}\cdot \bm{m}_{4}) \nonumber \\  
		& + & (\nabla_{r,1}\bm{m}_3)  \cdot (\mathcal{G}_{2,3}\cdot \bm{m}_{2} + \mathcal{G}_{3,4}\cdot \bm{m}_{4}) \nonumber \\  
		& - & \chi_3 \nabla_{r,1}(\mathcal{G}_{2,3} \cdot  \bm{m}_2 + \mathcal{G}_{3,4} \cdot  \bm{m}_4)  \cdot (\mathcal{G}_{2,3}\cdot \bm{m}_{2} + \mathcal{G}_{3,4}\cdot \bm{m}_{4}) \nonumber \\    
		& + & (\nabla_{r,1}\bm{m}_4)  \cdot (\mathcal{G}_{2,4}\cdot \bm{m}_{2} + \mathcal{G}_{3,4}\cdot \bm{m}_{3}) \nonumber \\  
		& - & \chi_4 \nabla_{r,1}(\mathcal{G}_{2,4} \cdot  \bm{m}_2 + \mathcal{G}_{3,4} \cdot  \bm{m}_3)  \cdot (\mathcal{G}_{2,4}\cdot \bm{m}_{2} + \mathcal{G}_{3,4}\cdot \bm{m}_{3}) \nonumber \\
	\end{eqnarray}
	\noindent 
	Reorganizing, we have
	\begin{eqnarray}
		\label{eq:4f1_11}
		\frac{\bm{f}_1}{\mu_0} & = &  \nabla_{r,1}  [ \underline{ \mathcal{G}_{1,2} : \bm{m}_2  \bm{m}_1 +  \mathcal{G}_{1,3} : \bm{m}_3  \bm{m}_1  +  \mathcal{G}_{1,4} : \bm{m}_4  \bm{m}_1 } \nonumber \\
		& + &  \underline{\mathcal{G}_{2,3} : \bm{m}_2  \bm{m}_3 + \mathcal{G}_{2,4} : \bm{m}_2  \bm{m}_4 + \mathcal{G}_{3,4} : \bm{m}_3  \bm{m}_4  }\nonumber \\
		& - &  \underline{\frac{\chi_2}{2}(\mathcal{G}_{1,2}\cdot \bm{m}_{1})^2 - \frac{\chi_1}{2}(\mathcal{G}_{1,2}\cdot \bm{m}_{2})^2 -\frac{\chi_3}{2}(\mathcal{G}_{1,3}\cdot \bm{m}_{1})^2 } \nonumber \\
		& - & \underline{ \frac{\chi_1}{2}(\mathcal{G}_{1,3}\cdot \bm{m}_{3})^2 - \frac{\chi_4}{2}(\mathcal{G}_{1,4}\cdot \bm{m}_{1})^2 - \frac{\chi_1}{2}(\mathcal{G}_{1,4}\cdot \bm{m}_{4})^2 } \nonumber \\
		& - & \underline{ \frac{\chi_2}{2}(\mathcal{G}_{2,3}\cdot \bm{m}_{3})^2 - \frac{\chi_2}{2}(\mathcal{G}_{2,4}\cdot \bm{m}_{4})^2 -\frac{\chi_3}{2}(\mathcal{G}_{2,3}\cdot \bm{m}_{2})^2  } \nonumber \\
		& - &  \underline{ \frac{\chi_3}{2}(\mathcal{G}_{3,4}\cdot \bm{m}_{4})^2  - \frac{\chi_4}{2}(\mathcal{G}_{2,4}\cdot \bm{m}_{2})^2 - \frac{\chi_4}{2}(\mathcal{G}_{3,4}\cdot \bm{m}_{3})^2  } \nonumber \\
		& - & \underline{ \chi_2 (\mathcal{G}_{2,3}\cdot \bm{m}_{3} \cdot \mathcal{G}_{1,2} \cdot  \bm{m}_1 ) - \chi_2(\mathcal{G}_{2,4}\cdot \bm{m}_{4} \cdot \mathcal{G}_{1,2} \cdot  \bm{m}_1) } \nonumber \\  
		& - & \underline{ \chi_1 (\mathcal{G}_{1,3}\cdot \bm{m}_{3} \cdot \mathcal{G}_{1,2} \cdot  \bm{m}_2 ) - \chi_1(\mathcal{G}_{1,4}\cdot \bm{m}_{4} \cdot \mathcal{G}_{1,2} \cdot  \bm{m}_2)  } \nonumber \\  
		& - & \underline{ \chi_3 (\mathcal{G}_{2,3}\cdot \bm{m}_{2} \cdot \mathcal{G}_{1,3} \cdot  \bm{m}_1 ) - \chi_3(\mathcal{G}_{3,4}\cdot \bm{m}_{4} \cdot \mathcal{G}_{1,3} \cdot  \bm{m}_1)  } \nonumber \\
		& - & \underline{ \chi_4 (\mathcal{G}_{2,4}\cdot \bm{m}_{2} \cdot \mathcal{G}_{1,4} \cdot  \bm{m}_1 ) - \chi_4(\mathcal{G}_{3,4}\cdot \bm{m}_{3} \cdot \mathcal{G}_{1,4} \cdot  \bm{m}_1) } \nonumber \\
		& - & \underline{ \chi_1 (\mathcal{G}_{1,3}\cdot \bm{m}_{3} \cdot \mathcal{G}_{1,4} \cdot  \bm{m}_4 ) } ] \nonumber \\    
		& - & \chi_2 \nabla_{r,1}(\mathcal{G}_{2,3} \cdot\bm{m}_3)\cdot(\mathcal{G}_{2,4} \cdot  \bm{m}_4) \nonumber \\
		& - & \chi_2 \nabla_{r,1}(\mathcal{G}_{2,4} \cdot \bm{m}_4)\cdot(\mathcal{G}_{2,3} \cdot  \bm{m}_3) \nonumber \\
		& - & \chi_3 \nabla_{r,1}(\mathcal{G}_{2,3} \cdot\bm{m}_2)\cdot(\mathcal{G}_{3,4} \cdot  \bm{m}_4) \nonumber \\
		& - & \chi_3 \nabla_{r,1}(\mathcal{G}_{3,4} \cdot \bm{m}_4)\cdot(\mathcal{G}_{2,3} \cdot  \bm{m}_2) \nonumber \\
		& - & \chi_4 \nabla_{r,1}(\mathcal{G}_{2,4} \cdot\bm{m}_2)\cdot(\mathcal{G}_{3,4} \cdot  \bm{m}_3) \nonumber \\
		& - & \chi_4 \nabla_{r,1}(\mathcal{G}_{3,4} \cdot \bm{m}_3)\cdot(\mathcal{G}_{2,4} \cdot  \bm{m}_2) \nonumber \\
	\end{eqnarray}
	\noindent 
	\newpage
	Using once again $\nabla(\bm{a}\cdot\bm{b}) = (\nabla\bm{a})\cdot\bm{b} + (\nabla\bm{b})\cdot\bm{a}$ to the appropriate terms, we are able to have all terms inside the gradient operator, then, the underline highlight is no longer necessary. 
	\begin{eqnarray}
		\label{eq:4f1_12}
		\frac{\bm{f}_1}{\mu_0} & = &  \nabla_{r,1}  [ \mathcal{G}_{1,2} : \bm{m}_2  \bm{m}_1 +  \mathcal{G}_{1,3} : \bm{m}_3  \bm{m}_1  +  \mathcal{G}_{1,4} : \bm{m}_4  \bm{m}_1  \nonumber \\
		& + &  \mathcal{G}_{2,3} : \bm{m}_2  \bm{m}_3 + \mathcal{G}_{2,4} : \bm{m}_2  \bm{m}_4 + \mathcal{G}_{3,4} : \bm{m}_3  \bm{m}_4  \nonumber \\
		& - &  \frac{\chi_2}{2}(\mathcal{G}_{1,2}\cdot \bm{m}_{1})^2 - \frac{\chi_1}{2}(\mathcal{G}_{1,2}\cdot \bm{m}_{2})^2 - \frac{\chi_3}{2}(\mathcal{G}_{1,3}\cdot \bm{m}_{1})^2 \nonumber \\
		& - &  \frac{\chi_1}{2}(\mathcal{G}_{1,3}\cdot \bm{m}_{3})^2  - \frac{\chi_4}{2}(\mathcal{G}_{1,4}\cdot \bm{m}_{1})^2 - \frac{\chi_1}{2}(\mathcal{G}_{1,4}\cdot \bm{m}_{4})^2 \nonumber \\
		& - &  \frac{\chi_2}{2}(\mathcal{G}_{2,3}\cdot \bm{m}_{3})^2 - \frac{\chi_2}{2}(\mathcal{G}_{2,4}\cdot \bm{m}_{4})^2  - \frac{\chi_3}{2}(\mathcal{G}_{2,3}\cdot \bm{m}_{2})^2 \nonumber \\
		& - &  \frac{\chi_3}{2}(\mathcal{G}_{3,4}\cdot \bm{m}_{4})^2 - \frac{\chi_4}{2}(\mathcal{G}_{2,4}\cdot \bm{m}_{2})^2 - \frac{\chi_4}{2}(\mathcal{G}_{3,4}\cdot \bm{m}_{3})^2  \nonumber \\
		& - & \chi_2 (\mathcal{G}_{2,3}\cdot \bm{m}_{3} \cdot \mathcal{G}_{1,2} \cdot  \bm{m}_1 ) - \chi_2(\mathcal{G}_{2,4}\cdot \bm{m}_{4} \cdot \mathcal{G}_{1,2} \cdot  \bm{m}_1) \nonumber \\  
		& - & \chi_1 (\mathcal{G}_{1,3}\cdot \bm{m}_{3} \cdot \mathcal{G}_{1,2} \cdot  \bm{m}_2 ) - \chi_1(\mathcal{G}_{1,4}\cdot \bm{m}_{4} \cdot \mathcal{G}_{1,2} \cdot  \bm{m}_2)  \nonumber \\  
		& - & \chi_3 (\mathcal{G}_{2,3}\cdot \bm{m}_{2} \cdot \mathcal{G}_{1,3} \cdot  \bm{m}_1 ) - \chi_3(\mathcal{G}_{3,4}\cdot \bm{m}_{4} \cdot \mathcal{G}_{1,3} \cdot  \bm{m}_1)  \nonumber \\
		& - & \chi_4 (\mathcal{G}_{2,4}\cdot \bm{m}_{2} \cdot \mathcal{G}_{1,4} \cdot  \bm{m}_1 ) - \chi_4(\mathcal{G}_{3,4}\cdot \bm{m}_{3} \cdot \mathcal{G}_{1,4} \cdot  \bm{m}_1) \nonumber \\
		& - & \chi_1 (\mathcal{G}_{1,3}\cdot \bm{m}_{3} \cdot \mathcal{G}_{1,4} \cdot  \bm{m}_4 ) - \chi_2 (\mathcal{G}_{2,3} \cdot\bm{m}_3\cdot\mathcal{G}_{2,4} \cdot  \bm{m}_4) \nonumber \\
		& - & \chi_3 (\mathcal{G}_{2,3} \cdot\bm{m}_2\cdot\mathcal{G}_{3,4} \cdot  \bm{m}_4) - \chi_4 (\mathcal{G}_{2,4} \cdot\bm{m}_2\cdot\mathcal{G}_{3,4} \cdot  \bm{m}_3)] \nonumber \\
	\end{eqnarray}
	\noindent 
	\newpage 
	
	Finally, the potential for 4 mutually interacting dipoles is writen as
	\begin{eqnarray}
		\label{eq:4U}
		\frac{U}{\mu_0} & = &  - \mathcal{G}_{1,2} : \bm{m}_2  \bm{m}_1 -  \mathcal{G}_{1,3} : \bm{m}_3  \bm{m}_1  -  \mathcal{G}_{1,4} : \bm{m}_4  \bm{m}_1  \nonumber \\
		& - &  \mathcal{G}_{2,3} : \bm{m}_2  \bm{m}_3 - \mathcal{G}_{2,4} : \bm{m}_2  \bm{m}_4 - \mathcal{G}_{3,4} : \bm{m}_3  \bm{m}_4  \nonumber \\
		& + &  \frac{\chi_2}{2}(\mathcal{G}_{1,2}\cdot \bm{m}_{1})^2 + \frac{\chi_1}{2}(\mathcal{G}_{1,2}\cdot \bm{m}_{2})^2 + \frac{\chi_3}{2}(\mathcal{G}_{1,3}\cdot \bm{m}_{1})^2\nonumber \\
		& + &  \frac{\chi_1}{2}(\mathcal{G}_{1,3}\cdot \bm{m}_{3})^2 + \frac{\chi_4}{2}(\mathcal{G}_{1,4}\cdot \bm{m}_{1})^2 + \frac{\chi_1}{2}(\mathcal{G}_{1,4}\cdot \bm{m}_{4})^2  \nonumber \\
		& + &  \frac{\chi_2}{2}(\mathcal{G}_{2,3}\cdot \bm{m}_{3})^2 + \frac{\chi_2}{2}(\mathcal{G}_{2,4}\cdot \bm{m}_{4})^2  + \frac{\chi_3}{2}(\mathcal{G}_{2,3}\cdot \bm{m}_{2})^2 \nonumber \\
		& + &  \frac{\chi_3}{2}(\mathcal{G}_{3,4}\cdot \bm{m}_{4})^2 + \frac{\chi_4}{2}(\mathcal{G}_{2,4}\cdot \bm{m}_{2})^2 + \frac{\chi_4}{2}(\mathcal{G}_{3,4}\cdot \bm{m}_{3})^2 \nonumber \\
		& + & \chi_2 (\mathcal{G}_{2,3}\cdot \bm{m}_{3} \cdot \mathcal{G}_{1,2} \cdot  \bm{m}_1 ) + \chi_2(\mathcal{G}_{2,4}\cdot \bm{m}_{4} \cdot \mathcal{G}_{1,2} \cdot  \bm{m}_1) \nonumber \\  
		& + & \chi_1 (\mathcal{G}_{1,3}\cdot \bm{m}_{3} \cdot \mathcal{G}_{1,2} \cdot  \bm{m}_2 ) + \chi_1(\mathcal{G}_{1,4}\cdot \bm{m}_{4} \cdot \mathcal{G}_{1,2} \cdot  \bm{m}_2)  \nonumber \\  
		& + & \chi_3 (\mathcal{G}_{2,3}\cdot \bm{m}_{2} \cdot \mathcal{G}_{1,3} \cdot  \bm{m}_1 ) + \chi_3(\mathcal{G}_{3,4}\cdot \bm{m}_{4} \cdot \mathcal{G}_{1,3} \cdot  \bm{m}_1)  \nonumber \\
		& + & \chi_4 (\mathcal{G}_{2,4}\cdot \bm{m}_{2} \cdot \mathcal{G}_{1,4} \cdot  \bm{m}_1 ) + \chi_4(\mathcal{G}_{3,4}\cdot \bm{m}_{3} \cdot \mathcal{G}_{1,4} \cdot  \bm{m}_1) \nonumber \\
		& + & \chi_1 (\mathcal{G}_{1,3}\cdot \bm{m}_{3} \cdot \mathcal{G}_{1,4} \cdot  \bm{m}_4 ) + \chi_2 (\mathcal{G}_{2,3} \cdot\bm{m}_3\cdot\mathcal{G}_{2,4} \cdot  \bm{m}_4) \nonumber \\
		& + & \chi_3 (\mathcal{G}_{2,3} \cdot\bm{m}_2\cdot\mathcal{G}_{3,4} \cdot  \bm{m}_4) + \chi_4 (\mathcal{G}_{2,4} \cdot\bm{m}_2\cdot\mathcal{G}_{3,4} \cdot  \bm{m}_3) \nonumber \\
	\end{eqnarray}
	\newpage
	
	\bibliography{apssamp}
	
\end{document}